\documentstyle[preprint,pre,epsf,aps]{revtex}
\begin{document}
\draft
\title {Exact Persistence Exponent for One-dimensional 
Potts Models with Parallel Dynamics}
\author{Gautam I. Menon\footnote{Email:menon@imsc.ernet.in}
and P. Ray\footnote{Email:ray@imsc.ernet.in}}
\address{The Institute of Mathematical Sciences, C.I.T. Campus, \\
Taramani, Chennai 600 113,\\
India}
\date{\today}
\maketitle

\begin{abstract}

We obtain $\theta_p(q) = 2\theta_s(q)$ for
one-dimensional $q$-state ferromagnetic Potts models
evolving under parallel dynamics at zero temperature
from an initially disordered state, where
$\theta_p(q)$ is the persistence exponent for parallel
dynamics and $\theta_s(q) = -\frac{1}{8}+
\frac{2}{\pi^2}[cos^{-1}\{(2-q)/q\sqrt{2}\}]^2$ [PRL,
{\bf 75}, 751, (1995)], the persistence exponent under
serial dynamics. This result is a consequence of an
exact, albeit non-trivial, mapping of the evolution of
configurations of Potts spins under parallel dynamics
to the dynamics of two decoupled reaction diffusion
systems.

\end{abstract}
\pacs{PACS:02.50.Ey, 05.40.-a, 05.10.Gg}

Consider a stochastic variable, such as the position
of a Brownian particle $x(t)$ moving in one dimension
as a function of time ($t$) or the state $\sigma(t)$~
($+1$~or~$-1$) of an Ising spin evolving under a
prescribed probabilistic dynamics.  The following
question arises naturally in many physical contexts:
What is the  probability that some attribute of the
system, such as the {\em state} of the Ising spin or,
for the Brownian particle, the {\em sign} of its
$x$-coordinate, is unchanged in an interval $[0,t]$?
Such questions relate to the ``persistence''
properties of stochastic processes and have attracted
both theoretical and experimental attention in recent
years\cite{derrida1,derrida,satya}.

This Letter studies persistence,
under {\em parallel} dynamics,
in one-dimensional ferromagnetic $q$-state
Potts models with nearest neighbor interactions,
described by the Hamiltonian
\begin{equation}
H = -\sum_i \delta_{\sigma_i,\sigma_{i+1}},
\end{equation}
where $\sigma_i$, the Potts spin at site $i$,
takes values $0 \ldots (q-1)$. In terms of the
conventional (serial) zero-temperature dynamics 
of this system, the problem of persistence 
is posed in the following way:
Start from initial configurations in which each spin takes
any of the allowed $q$ values with equal probability.
Then relax the system by allowing one spin at a time
to lower its energy by aligning itself with one of its
neighbours, chosen from right or left with equal
probability. Such a dynamics is serial (equivalently,
sequential)  since a single
spin update is performed at each step.  Now fix any
spin and ask: What is the probability, averaged
over initial configurations, that this spin
has {\em not} changed its state up to time $t$? This
quantity, the persistence probability $P(t)$, was 
first found numerically to decay as a power law, 
\begin{equation}
P(t) \sim \frac{1}{t^{\theta_s(q)}}, 
\label{persdef}
\end{equation}
with $\theta_s(q)$ a new, $q-$dependent non-trivial
{\em persistence} exponent; the subscript
denotes serial dynamics. A subsequent analytic {\it
tour de force} derived $\theta_s(q)$ exactly\cite{derrida},
\begin{equation} 
\theta_s(q) = -\frac{1}{8}+
\frac{2}{\pi^2}[cos^{-1}\{(2-q)/q\sqrt{2}\}]^2.
\label{persexact}
\end{equation} 
The persistence exponent is argued to be a {\em new}
exponent characterizing the dynamics\cite{satya}.

Barring this remarkable result, we know of very few
other {\em exact} results in the persistence problem
for many interacting variables. Further, these
results, besides being restricted to one dimensional
problems, have so far, with the exception of
Ref.\cite{ourpaper}, been confined to serial
dynamics.  The issue of the dynamics is important
here, since the persistence exponent does not share
the universal character of the static and dynamic
exponents. It is thus of obvious interest to find
other {\em exact} results for persistence in
interacting many-body systems, to understand the
origins of such scale invariant behaviour.

This Letter obtains the following result
connecting the persistence exponent under parallel
dynamics $\theta_p(q)$ for one-dimensional $q$-state
ferromagnetic Potts models with the corresponding
exponent $\theta_s(q)$ for serial dynamics:
\begin{equation} 
\theta_p(q) = 2\theta_s(q),
\label{conjecture}
\end{equation}
where $P(t) \sim 1/t^{\theta_p(q)}$ in the parallel 
case.  
We demonstrate the
validity of this result first numerically and then
provide a proof 
via a sequence of mappings to the dynamics of
the following reaction diffusion system:
\begin{eqnarray}
A + A & \rightarrow  \phi  &~~~~~~ {\rm Probability}~~1/(q-1) \nonumber \\ 
      & \rightarrow  A     &~~~~~~ {\rm Probability}~~(q-2)/(q-1) \nonumber \\
B + B & \rightarrow  \phi  &~~~~~~ {\rm Probability}~~1/(q-1) \nonumber \\
      & \rightarrow  B     &~~~~~~ {\rm Probability}~~(q-2)/(q-1),
\label{diffann}
\end{eqnarray}
describing particles of two species, labelled $A$ and
$B$, which diffuse and interact in one dimension.
Apart from the doubling of species, this reaction
scheme is identical to the one used in a domain wall
mapping for persistence in the Potts model with {\em
serial} dynamics, with the particles representing the
domain walls. The initial distribution of $A$ and $B$
particles can be assigned given any initial
configuration of Potts spins through the mapping we
describe below.  However, as we show here, the $A$ and
$B$ particles which enter the analysis of the case of
parallel dynamics are not the same as the domain walls
in the serial case.

In terms of this mapping, whether a given site is
persistent or not at time $t$ is given by 
the probability that it is not crossed by either a
particle of type A {\em or} B upto time $t$.
Since the two reactions are
independent, provided there are no significant
correlations in the initial placement of A and B
particles, the {\em joint} probability that a given
site is persistent with respect to the motion of both
A and B particles simply factors into the product of
independent probabilities at long times, as 
\begin{equation} 
P(t) \sim \frac{1}{t^{\theta_s(q)}}\frac{1}{t^{\theta_s(q)}}
\sim \frac{1}{t^{2\theta_s(q)}} \sim
\frac{1}{t^{\theta_p(q)}}, 
\label{decoupling}
\end{equation} 
yielding immediately the result of Eq.~\ref{conjecture}.

We first present results from our numerical
simulations of this problem.  Each spin can take one
of~$(q-1)$~values, which we take, for concreteness, to
be represented by the integers $\{0 \ldots (q-1)\}$.  The
zero temperature dynamics evolves a spin configuration
$\{\sigma(t)\}$ at time $t$ to a configuration
$\{\sigma(t+1)\}$ at time $t+1$ through the following
simple rule:  Each spin at time $t+1$ assumes the
value of one of its neighboring spins at time $t$,
chosen from right or left with equal probability.
Each such step in time constitutes a single Monte
Carlo step. The parallel nature of the dynamics
follows from the fact that {\em all} spins are updated
together in a single time step.

We have simulated parallel dynamics using the above
rules on Potts systems with different $q$ values, on
lattices of linear size $L=10^5$ sites and for times
$t \leq 10^4$, applying periodic boundary conditions.
We average over a fairly large number of initial
conditions, typically $10^2-10^3$, starting from
configurations in which each spin is independently
assigned a value in the range $\{0 \ldots (q-1)\}$
with equal probability.  We compute the standard
persistence probability $P(t)$, defined as the
probability that the spin at a given site has not
changed its state up to time $t$, averaged over all
sites and over an ensemble of initial conditions.

Fig. 1 shows the persistence probability $P(t)$ for
ferromagnetic Potts systems, evolving under parallel
dynamics, for $q$ = 3, 5, 8 and 20. Given Eqs.
\ref{persexact} and \ref{conjecture}, we expect a
power law tail to $P(t)$ with exponents $\theta_p(3)
\simeq 1.076,~\theta_p(5) \simeq 1.386,~\theta_p(8)
\simeq 1.588$~and $\theta_p(20)~\simeq 1.821$. As can
be seen from the fits displayed, the data are
consistent with our proposal.

In a recent paper, coauthored with
Shukla\cite{ourpaper}, we studied the Ising version of
the problem of persistence with parallel dynamics,
obtaining the result of Eq.~\ref{conjecture}
specialized to $q=2$ {\it i.e.} $\theta_p(2) = 3/4 =
2\theta_s(2) = 2(3/8)$. We briefly sketch the mapping
which enabled us to show this exactly and then
describe how such a mapping is extended to solve the
more general Potts case.

For the one-dimensional Ising model with serial
dynamics, domain walls obey the reaction diffusion
scheme $A+A \rightarrow 0$. Whether a site is
persistent or not at some late time $t$, is equivalent
to asking whether it has been crossed by a domain wall
at an earlier time.  As shown in Ref\cite{ourpaper},
the diffusing particles whose motion and annihilation
govern the persistence properties of the Ising model
with {\em parallel} dynamics are {\em not} the domain
walls of the serial case, but entities which we termed
as {\em zero-field} sites {\it i.e.} sites in which
the spin is aligned with either one, but not both, of
its neighbours. 

Given any configuration of the Ising
model, all sites can be classified (uniquely) as
stable (S), unstable (U) or zero-field (Z) depending
on whether they will flip with probability $0,1$ or
$1/2$ respectively in the next time step.  While pairs
of {\em permanently bound} zero-field sites can be
identified with the domain walls of the serial case,
the two sites constituting such a pair can move
independently under parallel dynamics. A region of
stable sites is always bounded by zero-field sites; a
site embedded in an initially stable region is
persistent upto time $t$ if it has not been crossed by
a zero-field site in the interval $[0,t]$.  The
doubling of the persistence exponent is a consequence
of the fact that zero-field sites on A and B
sublattices decouple from each other under the
dynamics. Identifying zero-field sites on the $A$ (or
$B$) sublattice as particles labelled $A$ or $B$
respectively, the following equations describe the
interaction of these particles:  $A+A \rightarrow 0$
and $B+B \rightarrow 0$; these equations can be
obtained from Eq.~\ref{diffann} upon setting $q$ to
2.  As shown in Ref.\cite{ourpaper}, zero-field sites
on the A(B) sublattice at time $t$ hop to sites on the
B(A) sublattice at time $t+1$. The $A$ and $B$ labels
are identified with respect to an arbitrary initial
time, chosen for convenience to be $t=0$.

It is tempting to conjecture that a similar sublattice
decoupling is responsible for the doubling of the
persistence exponent with parallel dynamics {\it vis.
a vis}~the serial case, for general $q$. This idea, as
we will show, is essentially correct, but for the
following caveat:  The A and B particles are {\em not}
simply the zero-field sites of the Potts model. This
result follows from the observation that the
classification of all spins as stable, unstable or
zero-field, as outlined in the Ising case, is {\em
incomplete} when applied to the problem of persistence
in Potts models with $q>2$.

Arguments against a naive generalization of ideas from
the Ising case follow from inspecting the pictures of
Potts configurations evolving under parallel dynamics
shown in Figs. 2(a) and 2(b). In these pictures,
obtained from simulations of a 3-state Potts model,
the configuration corresponding to the earliest time
is shown on the top row, while each subsequent row
represents the configuration at the following time
step.  We have identified and labelled zero-field
sites in terms of the $A$ and $B$ particles described
above. Note that while A(B) particles hop at each time
step to sites on the B(A) sublattice in most of the
updates shown, this strict independance of the two
sublattices is violated in the penultimate row of
Fig.  2(a).  Evidently, the dynamics of the zero-field
sites alone, unlike the case for the one-dimensional
Ising model, does not decouple the two sublattices in
the more general Potts case.  A related difficulty is
apparent from inspection of the configurations of Fig.
2(b). Note that an $AB$ pair appears to be {\em
created} upon evolution.  Such a reaction cannot be
described by Eq.~\ref{diffann}, which yields a
strictly {\em non-increasing} number of $A$ and $B$
particles.

Our mapping to the reaction diffusion process
described by Eq.~\ref{diffann} in terms of
configurations of the $q-$state Potts model proceeds
as follows: For the purposes of illustration, we
consider a 3-state Potts model in which the states are
labelled by 0, 1 and 2.  If the spin at site $i$ is in
a particular Potts state $q$ while its neighbours are
in the {\em same} Potts state, for example the central
spin in the sequence $\ldots 000 \ldots$, that site is
labelled a {\em stable} ($S$) site. If a given spin at
site $i$ has one spin (on either side) in the same
state, while the other spin is in a  {\em different}
state, for example the central spin in $\ldots 002
\ldots$, we will term such a spin as a zero-field
($Z$) site. As in the Ising case, we will label these
zero-field sites as A or B, depending on the
sublattice to which they can be assigned.  Note that a
region of stable sites is bordered by zero-field
sites. 

The feature which makes this problem non-trivially
different from the Ising case is the following: We
must distinguish {\em two} categories of unstable
sites:  unstable sites of type I, labelled U$_1$, have
both neighbours in the {\em same} state which,
however, differs from the state of the  central spin.
For example, the central spin in the configuration
$\ldots 202\ldots$, is an unstable spin of type I.
Finally, we will define unstable sites of type II,
labelled U$_2$, in which the spins neighbouring a
central spin belong to Potts states which differ from
each other as well as from the state of the central
spin, the central spin in the configuration $ \ldots
012 \ldots$ being an example.  Note that unstable
sites of type II are absent in the Ising model.

We have now provided a prescription for mapping, given
the rules above, any configuration of Potts spins to
configurations of S,Z, U$_1$ and U$_2$. In the Ising
case, U$_2$ sites being absent, we proceeded to
identify the zero-field sites ($Z$) as particles of
type A or B and went on to show that this mapping
yielded precisely Eq.~\ref{diffann} above in the $q=2$
limit, as a consequence of the decoupling of the two
sublattices.  The central idea which allows us to
extend this intuition to the Potts model is the
following:  We {\em relabel} unstable sites of type
II, {\it i.e.} sites labelled U$_2$, as A or B
particles, depending on the sublattice to which they
belong at $t=0$. Given this relabelling, it is now a
trivial, though tedious, exercise to verify that the
rules for the interaction of A and B particles are
precisely those defined by Eq.~\ref{diffann} and that
the decoupling of the dynamics of A and B particles
follows exactly.  The chief result of this paper, our
Eq.~\ref{conjecture}, is then a natural consequence of
the arguments which led to Eq.~\ref{decoupling}.

To make this remarkable simplification explicit, we
have taken the configurations of Figs. 2(a) and (b)
and added the new $A$ and $B$ particles obtained  by
appropriately labelling unstable sites of type II  as
described above. These configurations are shown in
Figs 3(a) and (b). Note that the incorporation of
these new sites immediately solves both problems with
the earlier mapping (Figs. 2(a) and (b)) which used
only the zero-field sites:  The configuration of Fig.
2(a) which showed an A particle remaining on the same
sublattice at the following time step is now
interpreted in terms of the hopping of both the $A$
particle and an adjacent $B$ particle associated with
an unstable site of type II.  The difficulty of the
apparent {\em creation} of an $AB$ pair (Fig. 2(b)),
is removed by recognizing that this simply arises due
to the interconversion of a pair of unstable sites of
type II and a pair of zero-field sites.

We now demonstrate, for a simple case, how the rates
of Eq.~\ref{diffann} are obtained as a consequence of
this mapping.  Consider the time evolution of the
following spin configuration in a Potts model under
parallel dynamics:~$aaaabcccc$, where Potts states $a$
and $c$ form stable domains.  If $b$ is distinct from
both $a$ and $c$ (with $a \neq c$), then the mapping
described above leads to the following description of
the configuration:~$.......ABA.....$, where the first
zero-field site has been taken to belong, arbitrarily,
to sublattice $A$. Note that the two $A$ particles are
associated with zero-field sites, whereas the $B$
particle is associated with an unstable site of type
II.  The following are the possible configurations at
the next time step $aaaabcccc \rightarrow aaaaacccc,
aaaacbccc, aaaaabccc, aaaaccccc, aaabacccc, aaabccccc,
aaababccc, aaabcbccc$.  For distinct $a,b$ and $c$,
fixing $a$ would give $(q-1)$ choices for $c$ and
$(q-2)$ possibilities for $b$, where $q$ is the number
of Potts states, yielding a total of~$(q-1)(q-2)$
possible variations on b and c, given a.

Now notice that precisely two of these final states,
$aaaaacccc$ and $aaaaccccc$, correspond to the
reaction $A+A \rightarrow A$, with the $A$ particle
transforming into a B particle at the next time step;
see Fig.~3 for an illustration. All other states can
be interpreted in terms of the diffusion of the $A$
and $B$ particles to neighbour sites, keeping in mind
the interchange of A and B labels at succeeding time
steps.  Now if $a = c$ while $b \neq a,c$, as in the
initial configuration $aaaabaaaaa$, fixing $a$ would give
$(q-1)$ choices for $b$.  However, if $a=c$, the spin
flip at the ``b'' site yields, for the final
configurations corresponding to the reaction $A+A
\rightarrow A$ above, the reaction $aaaabaaaaa
\rightarrow aaaaaaaaaa$, interpreted as $A+A
\rightarrow \phi$ in terms of our mapping.

Starting from an initial disordered configuration and
averaging over all initial configurations, one would
expect configurations of the type ($aaaabcccc$), with
$a \neq b \neq c$, and ($aaaabaaaa$) with $a \neq b$,
to appear in the relative ratio $(q-2)(q-1):(q-1)$. In
terms of the dynamics of the $A$ and $B$ particles,
this translates into the reaction $A+A \rightarrow A$
occuring with probability $(q-2)/(q-1)$ {\it vis a
vis} the reaction $A+A \rightarrow \phi$ which occurs
with probability $1/(q-1)$. This justifies the rates
which appear in our mapping to the system of decoupled
reaction diffusion equations illustrated in
Eq.~\ref{diffann}.

A straightforward corollary to these results is the
following: The persistence of a given spin, measured
through configurations separated by {\em two} time
steps, should decay as $P(t) \sim
1/t^{\theta_s(q)}$~{\it i.e.} with the serial exponent
rather than the parallel one. This is a simple
consequence of the fact that only {\em one} of the
sublattices contributes to persistence in this case.
This exact result is interesting in view of recent
work on persistence with {\em discrete sampling}, in
which an underlying continuous time (as opposed to
discrete in the case studied here) process is sampled
at regular time intervals and its persistence
properties studied\cite{discrete,discrete1}.

In conclusion, we have described a mapping between
configurations of a one-dimensional $q-$state
ferromagnetic Potts model evolving under parallel
dynamics and a system of two {\em decoupled}
reaction-diffusion equations representing the motion
and interaction of the boundary sites of stable
domains. Using this mapping, we derive a general
result connecting persistence exponents obtained with
serial and parallel dynamics for this model. This
mapping relied on identifying the crucial and somewhat
unexpected role of what we have called unstable sites
of type II.  While the mapping here addressed the
specific problem of the persistence exponent, it may
be of more general use in the computation of other
statistical properties of Potts models evolving under
parallel dynamics.

We thank P. Shukla for useful dicussions in the
context of Ref~\cite{ourpaper}.

\begin{figure}[htp]
\vskip 2truecm
{\includegraphics{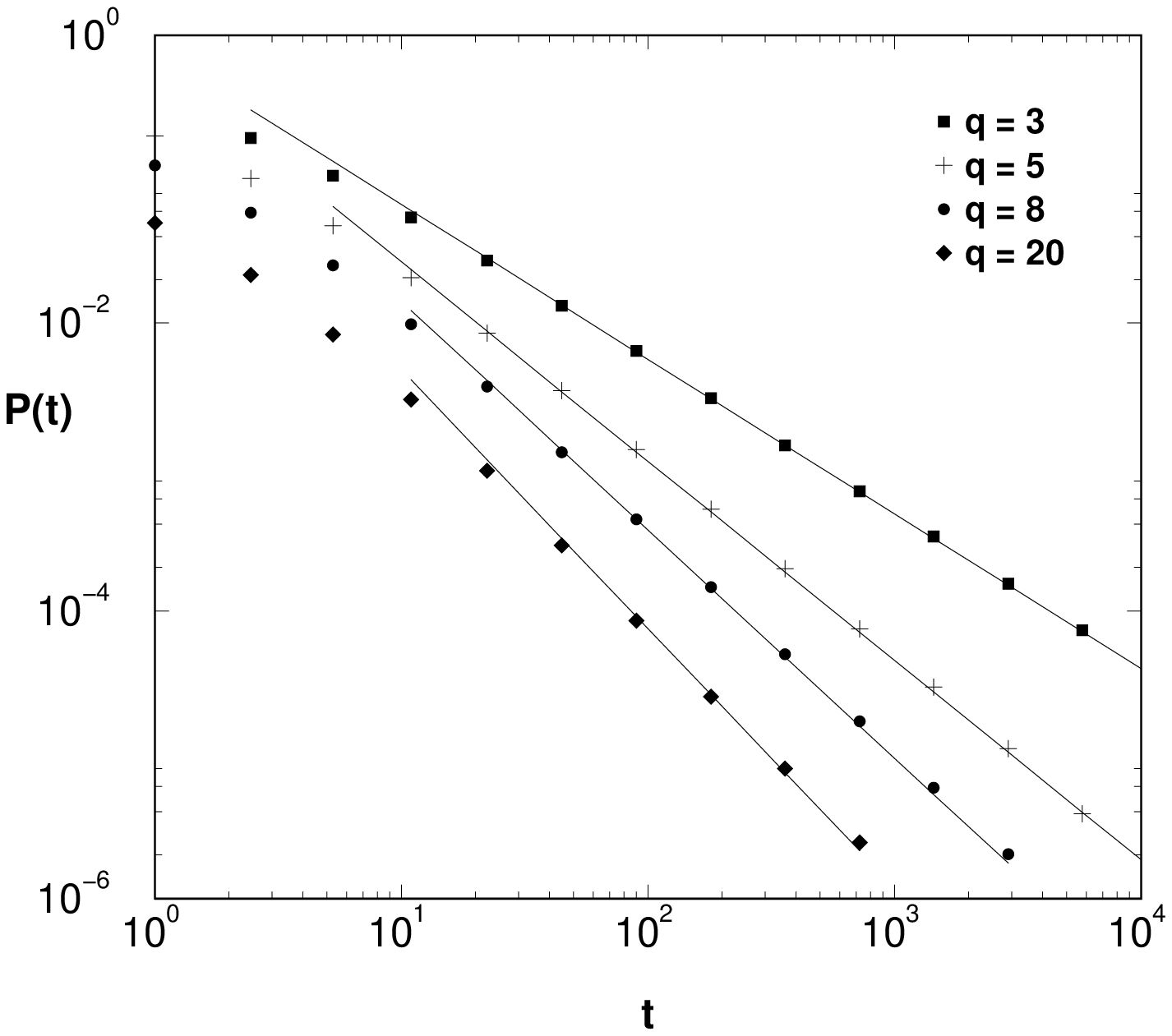}}
\vskip -1truecm
\caption
{
The persistence probability $P(t)$ in 1d
ferromagnetic $q-$ state Potts models evolving from
random configurations, plotted against time
$t$ in a logarithmic scale, for 
q = $3, 5,8$ and $20$. The dynamics is parallel;
{\em all} spins are updated in a single time step.
The lines represent fits with exponents $\theta_p(3)
\simeq 1.076,~\theta_p(5) \simeq 1.386,~\theta_p(8)
\simeq 1.588$~and $\theta_p(20)~\simeq 1.821$ [see
text].
}
\label{Fig1}
\end{figure}

\newpage

\begin{figure}[htp]
\vskip2truecm
{\includegraphics{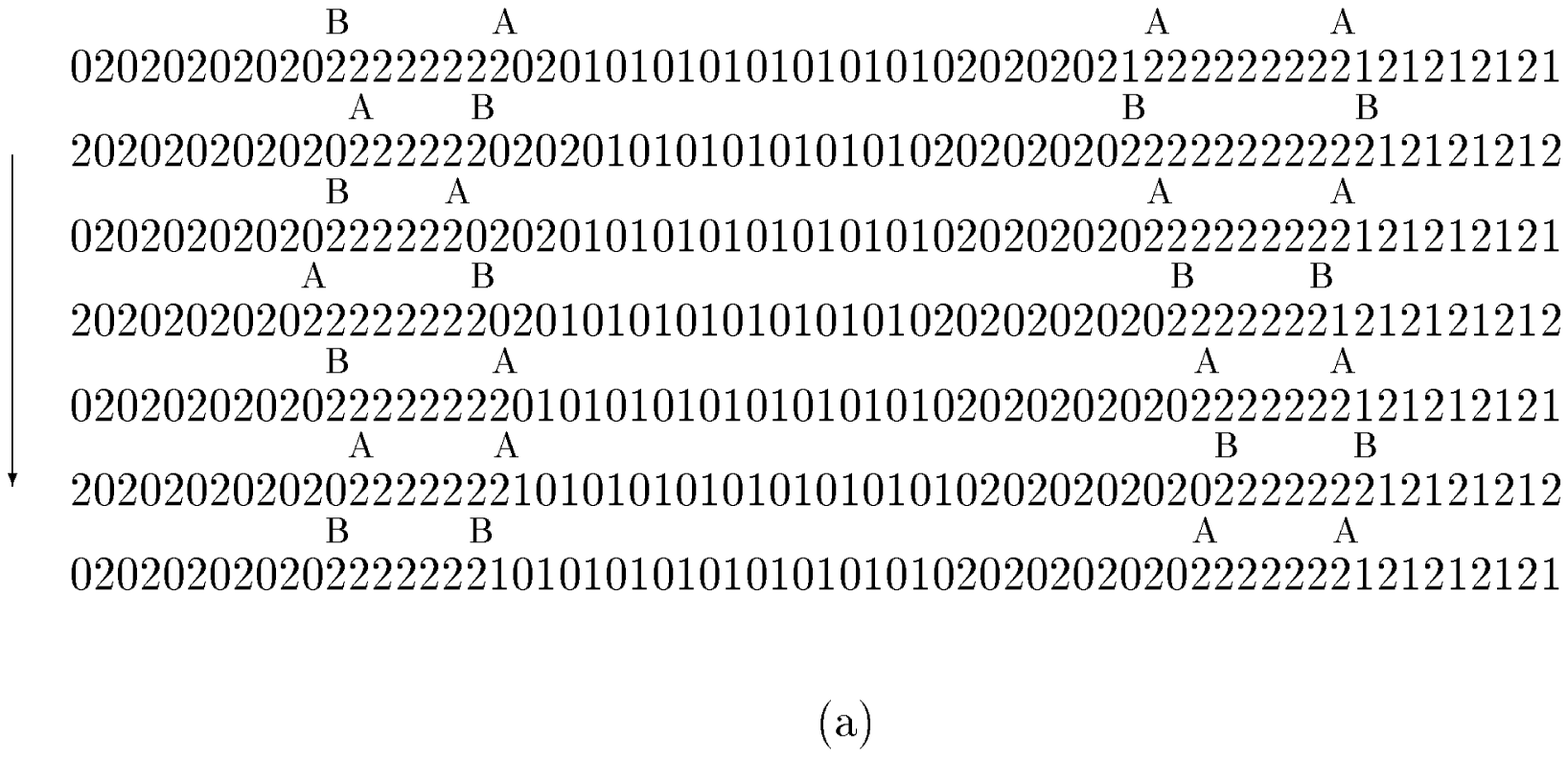}}
{\includegraphics{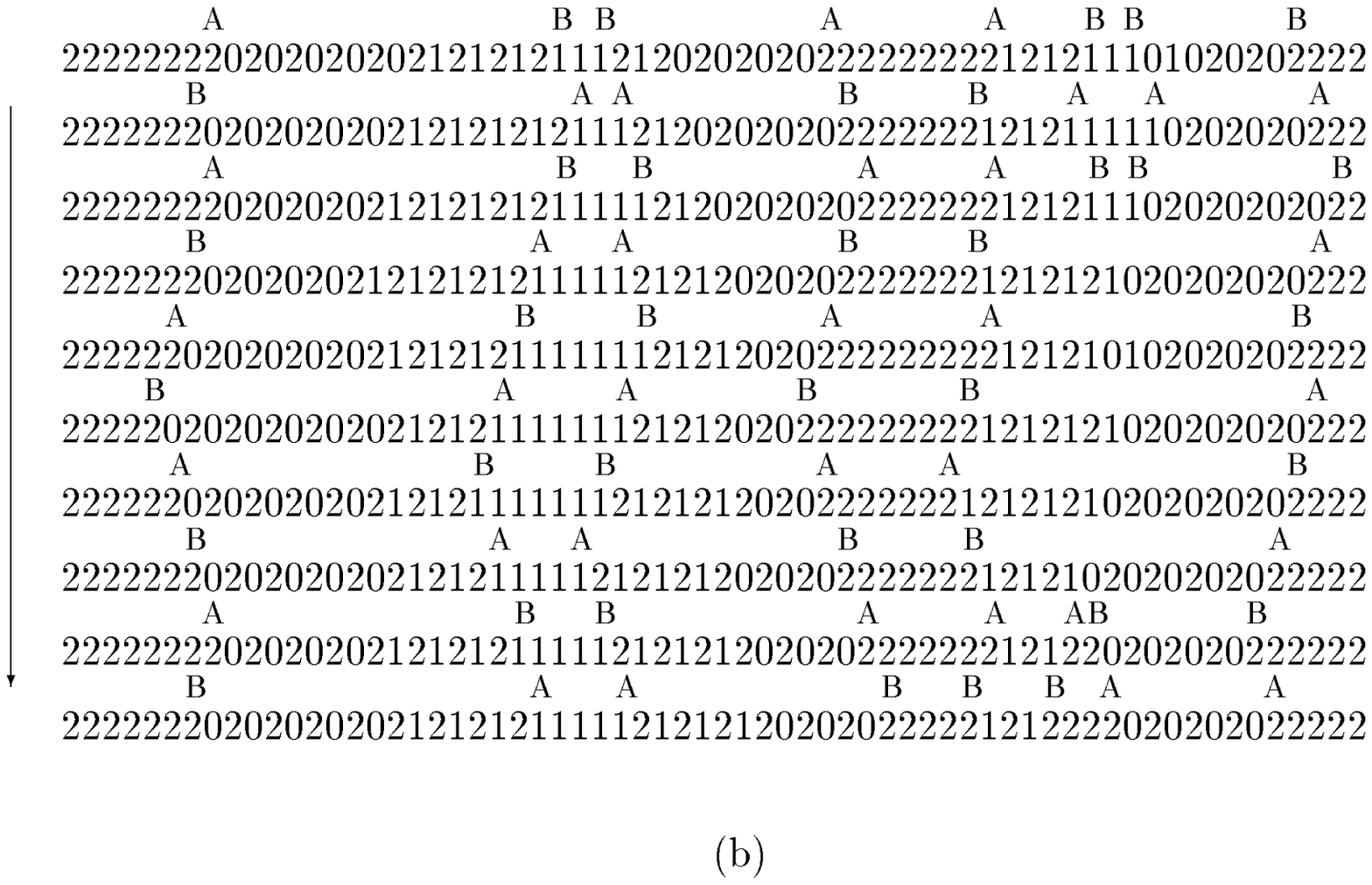}}
\vskip -3truecm
\caption
{
Evolution of configurations in a 3-state Potts model,
in which only zero-field sites (see text) are
identified and labelled by the sublattice to which
they belong.  The configurations are separated by a
time step; earlier times appear to the {\em top} of
the figure. Fig. 2(a) illustrates how the alternation  of
sublattices at succeeding time steps appears to be
violated while 2(b) illustrates how zero-field sites
can be destroyed as well as {\em created}, in contrast
to the results of a similar mapping for the Ising
model.
}
\label{Fig2}
\end{figure}

\newpage

\begin{figure}[htp]
\vskip2truecm
{\includegraphics{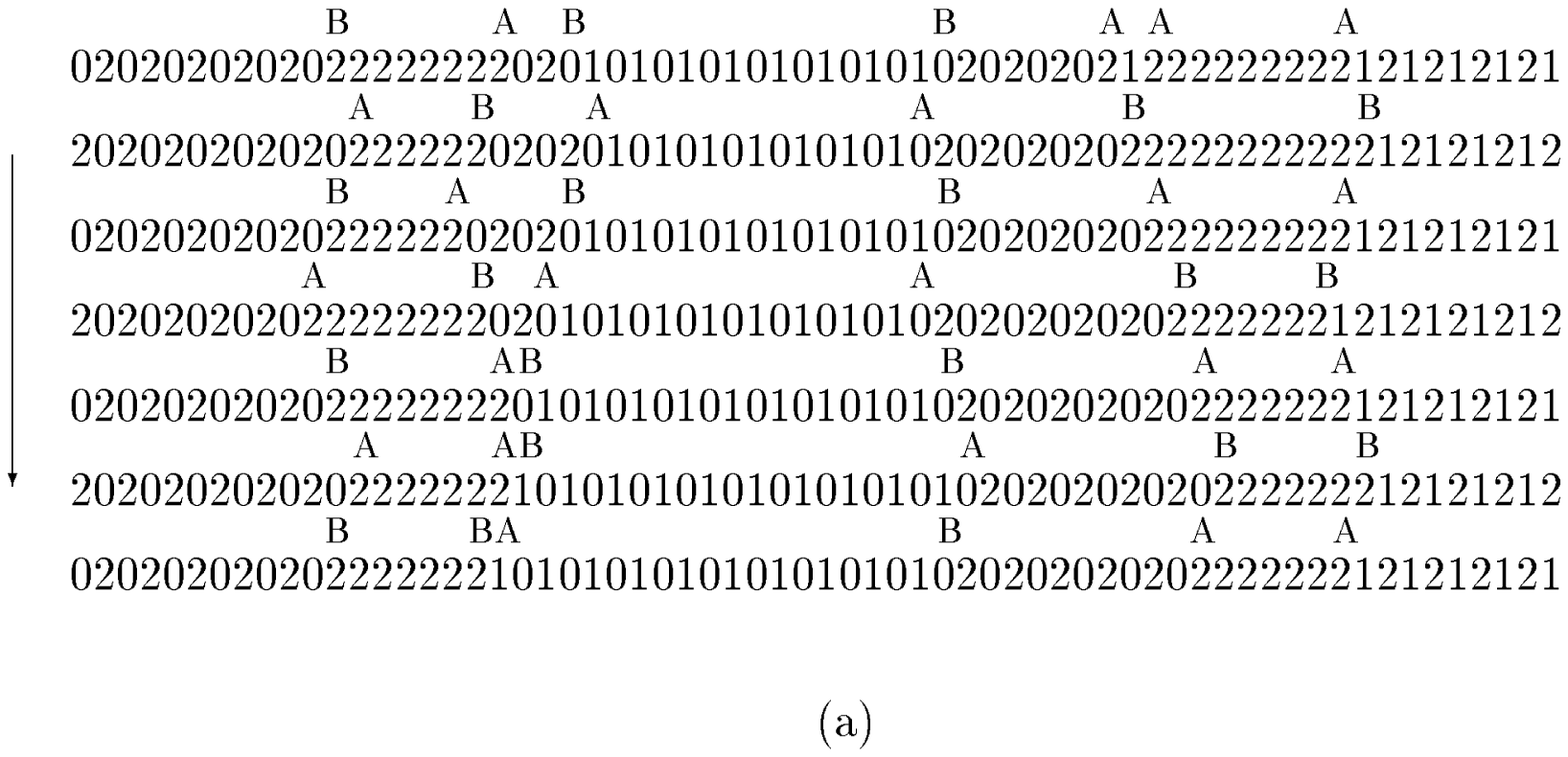}}
{\includegraphics{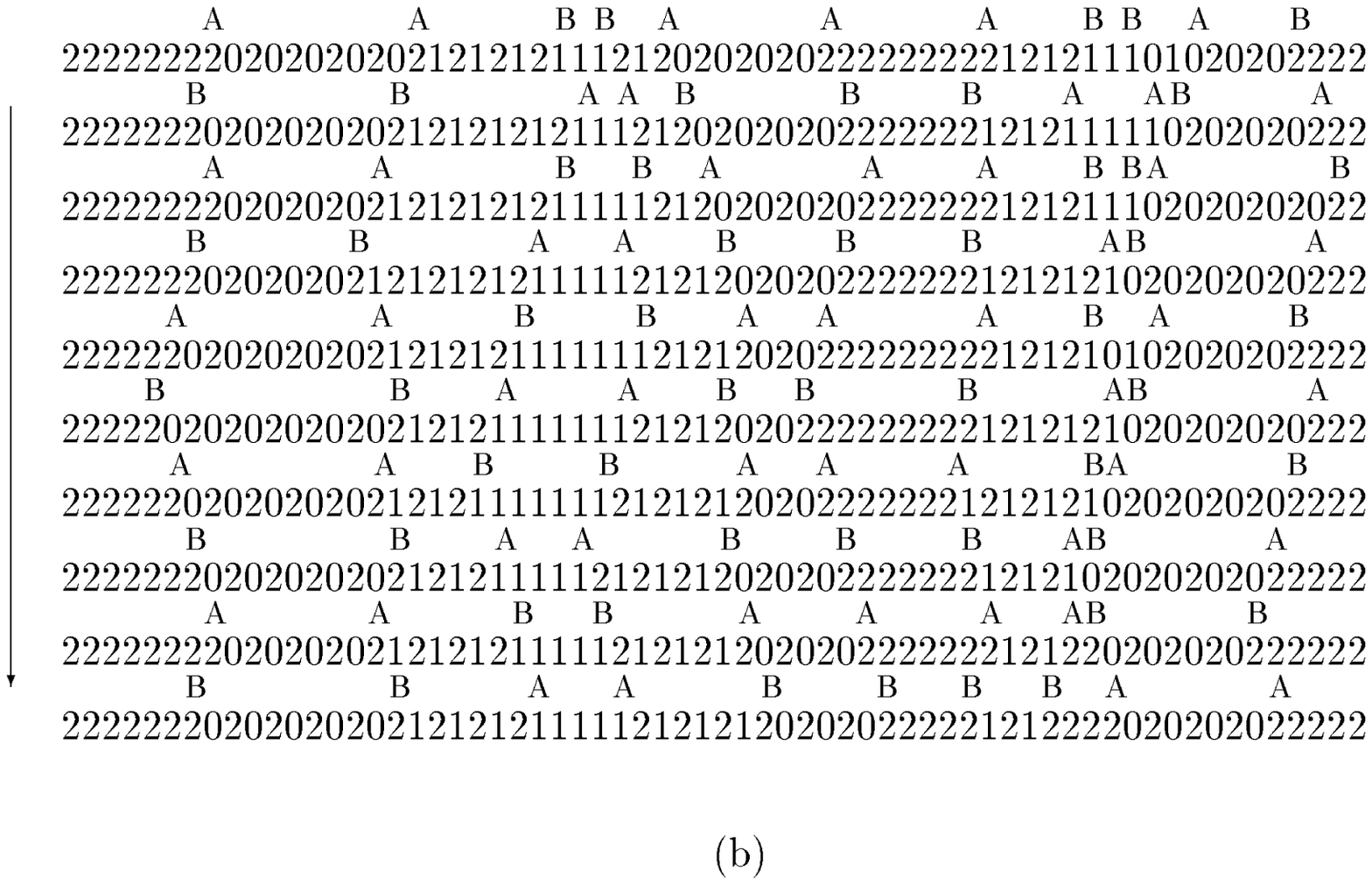}}
\vskip -1truecm
\caption
{
Evolution of the configurations of the 3-state Potts
model illustrated in Fig 2, in which unstable sites of
type-II (see text) in addition to zero-field sites are
identified and labelled according to the sublattice on
which they are present.  The configurations are
separated by a time step; earlier times appear to the
{\em top} of the figure.  The reaction scheme
discussed in the text is precisely followed upon the
inclusion of unstable sites of type II.  Note that the
two-sublattice decoupling is an exact feature of the
model in which both zero-field sites as well as
unstable sites of type-II are identified with the
diffusing and annihilating random walkers of the
reaction diffusion scheme outlined in the text.
}
\label{Fig3}
\end{figure}

\end{document}